\documentclass[twocolumn, pra, showpacs,superscriptaddress]{revtex4}
\usepackage{amsfonts}
\usepackage{amsmath}
\usepackage{amssymb}
\usepackage{graphicx}

\begin{document}

\author{Jie Zhang$^{\ast}$}
\affiliation{College of Physics and optoelectronics, Taiyuan University of Technology,
Taiyuan 030024, Shanxi, China}
\title{Quaternary-singlet\ State of Spin-1 Bosons in Optical Lattice}

\begin{abstract}
We present the quantum ground state properties of $^{23}$Na spinor
condensates, which is confined in a periodic or double-well potential and
subject to a magnetic dipole-dipole interaction between nearby wells. A
novel singlet state arise in the system and can be discussed in explicit
form. Caused by the competition between the intra-site spin exchange
interactions and the inter-site dipole-dipole interactions, this
quaternary-singlet\ state is a entangled state formed by at lest four
particles and vanish the total spin. This is distinct from the direct
product of the two conventional singlet pairs.
\end{abstract}

\maketitle

\section{\protect\bigskip Introduction}

The Heisenberg model of spin-spin interactions defined by$H=J\sum_{<i>j>}%
\mathbf{F}_{i}\cdot \mathbf{F}_{j}$ is often considered as the starting
point for understanding many complex magnetic structures in solids such as
ferromagnetism and antiferromagnetism at temperatures below the Curie
temperature. This Hamiltonian arises from the direct Coulomb interaction
among electrons and the Pauli exclusion principle,with $\mathbf{F}_{i}$ the
spin operator for the $i$th electron.

The study of bosonic spin-spin interactions rise since the success in
trapping a $^{23}$Na condensate in an optical potential \cite{D. M.
Stamper-Kurn}, where spin degrees of freedom are liberated and it give rise
to a rich variety phenomena such as spin domain formations \cite{Stenger},
spin mixing dynamics \cite{MSChang}, topological defects and so on. The
properties of such a three-component spinor condensate were first studied
with atomic spin coupling interaction takes the form $V(\mathbf{r}%
)=(c_{0}+c_{2}\mathbf{F}_{1}\cdot \mathbf{F}_{2})\delta (\mathbf{r})$ by Ho%
\cite{Ho} and Ohmi \cite{Ohmi}, and had been implemented experimentally \cite%
{Stenger}, where two different spin-dependent phases exist: the so-called
antiferromagnetic and ferromagnetic states for $^{23}$Na and $^{87}$Rb
atomic condensates respectively.

This spinor BECs can be confined in optical lattices, which offers a unique
opportunity to study magnetic properties of matter with tunable
parameters.The quantum phase transition from the superfluid phase to the
Mott insulating state is well described by the spinless Bose-Hubbard
Hamiltonian\cite{BH,BH2} and was demonstrated in experiments\cite{BHEX,BHEX2}%
.With the atomic spin coupling interaction involved, the Bose-Hubbard
Hamiltonian is studied in the Mott insulating regime, where phase coherence
or superfluidity is lost, and atoms are localized with number fluctuations
suppressed. Such a insulating state represents a correlated many-body state
of bosons and the calculations of the magnetic properties within
spin-exchange interactions have been carried out sufficiently\cite%
{before1,before2,before3,before4,before5,before6,before7}.

In addition to the spin-exchange interaction, there exists another important
type of magnetic interaction, the magnetic dipole-dipole interaction\cite%
{dipole1,dipole2,dipole3,dipole4,dipole5,dipole6,dipole7,dipole8}, which
plays an important role in domain formation in macroscopic samples. This
model includes the long-range magnetic dipole-dipole interaction between
different lattice sites, but neglects it within each site\cite{before4},
assuming that it is much weaker than the s-wave interaction described.

It is predicted that the ground state of $^{23}$Na BEC ($c_{2}>0$) is a spin
singlet with properties ($n_{1}=n_{0}=n_{-1}=N/3$) \cite{C. K. Law} contrast
with those of mean field prediction. Soon, Ho and Yip \cite{Ho and Yip} show
that this spin singlet state is a fragmented condensate with anomalously
large number fluctuations and thus has fragile stability. The remarkable
nature of this fragmentation is that the single particle reduced density
matrix gives three macroscopic eigenvalues (above) with large number
fluctuations $\Delta n_{1,0,-1}\sim N$. In this paper, we consider
specifically the case of spinor $^{23}$Na condensates and discuss the ground
state properties and quantum number fluctuations in the Zeeman components.

\section{The model Hamiltonian}

The simple two well model Hamiltonian includes $\hat{H}=\hat{H}_{1}+\hat{H}%
_{2}+\hat{H}_{12}$ where

\begin{eqnarray}
\hat{H}_{1} &=&\int d\mathbf{r}\left\{ \hat{\Psi}_{\alpha }^{\dag }(\frac{%
\hbar ^{2}}{2M_{1}}\nabla ^{2}+V_{trap})\hat{\Psi}_{\alpha }+\frac{\alpha
_{1}}{2}\hat{\Psi}_{\alpha }^{\dag }\hat{\Psi}_{\beta }^{\dag }\hat{\Psi}%
_{\beta }\hat{\Psi}_{\alpha }\right.  \notag \\
&&\left. +\frac{\beta _{1}}{2}\hat{\Psi}_{\alpha }^{\dag }\hat{\Psi}_{\beta
}^{\dag }\mathbf{F}_{\alpha \nu }\cdot \mathbf{F}_{\beta \mu }\hat{\Psi}%
_{\mu }\hat{\Psi}_{\nu }\right\}
\end{eqnarray}%
is the in-site spinor BEC Hamiltonian\cite{Ho,Ohmi}, the spin-dependent $%
\delta $ interaction is $V_{\delta }(\mathbf{r})=(\alpha +\beta \mathbf{F}%
\cdot \mathbf{F})\delta (\mathbf{r})$ with $\alpha ,\beta $ characterize the
short-range spin-independent and spin-changing s-wave collisions,
respectively. $\mathbf{\hat{F}}=\hat{a}_{\alpha }^{\dag }\mathbf{F}_{\alpha
\beta }\hat{a}_{\beta }$ is defined in terms of the $3\times 3$ spin-1
matrices $\mathbf{F}_{\alpha \beta }$, and $\alpha ,\beta =1,0,-1$ describe
the three Zeeman levels with repeated indices to be summed over. $H_{2}$ is
identical to $H_{1}$ except for the substitution of subscript $1$ by $2$ and
$\hat{\Psi}_{i}$ by $\hat{\Phi}_{i}$.

The dipole-dipole Hamiltonian is
\begin{equation}
\hat{H}_{12}=\int d\mathbf{r}\hat{\Psi}_{\alpha }^{\dag }\hat{\Phi}_{\beta
}^{\dag }V_{dd}(\mathbf{r}_{1}-\mathbf{r}_{2})\hat{\Phi}_{\beta }\hat{\Psi}%
_{\alpha }
\end{equation}%
with
\begin{equation}
V_{dd}=\frac{\mu _{0}}{4\pi }\left[ \frac{\mathbf{d}_{1}\cdot \mathbf{d}_{2}%
}{\left\vert \mathbf{r}_{12}\right\vert ^{3}}-\frac{3(\mathbf{d}_{1}\cdot
\mathbf{\hat{r}}_{12})(\mathbf{d}_{2}\cdot \mathbf{\hat{r}}_{12})}{%
\left\vert \mathbf{r}_{12}\right\vert ^{3}}\right]
\end{equation}%
$\mu _{0}$ the magnetic permeability of vacuum, $\mathbf{d}_{\mathbf{1,2}}%
\mathbf{=}g_{F}\mu _{B}\mathbf{F}_{1,2}$ with $g_{F}\mu _{B}$ the
gyromagnetic ratio, and $\mathbf{r}_{12}=$ $\mathbf{r}_{1}$ $-$ $\mathbf{r}%
_{2}$, $\mathbf{\hat{r}}_{12}$ $=\mathbf{r}_{12}/\left\vert \mathbf{r}%
_{12}\right\vert $, $\mathbf{r}_{1,2}$ is the coordinate of the $1,2$ site$.$

\bigskip We adopt the single model approximation \cite{C. K. Law,SMA1,SMA2}
for each of the two spinor condensates in the nearby sites with modes $\Psi (%
\mathbf{r})$ and $\Phi (\mathbf{r})$, i.e., setting
\begin{equation}
\hat{\Psi}_{i}=\hat{a}_{i}\Psi ,\text{\qquad }\hat{\Phi}_{i}=\hat{b}_{i}\Phi
,\text{ }i=1,0,-1
\end{equation}%
with $\hat{a}_{i}$ ($\hat{b}_{i}$) the annihilation operator for the
ferromagnetic (polar) atoms satisfying $\left[ \hat{a}_{i},\hat{a}_{j}\right]
=0$ and $\left[ \hat{a}_{i},\hat{a}_{j}^{\dag }\right] =\delta _{ij}$ (and
the same form of commutations for $\hat{b}_{i}$). substitute $\hat{\Psi}_{i},%
\hat{\Phi}_{i}$ into $\hat{H}_{1},\hat{H}_{2},$ The spin-independent part
can be reduced to a constant operator for a fix number of atoms $%
N=N_{1}+N_{0}+N_{-1},$

\begin{align}
(\frac{\hbar ^{2}}{2M_{1}}\nabla ^{2}+U_{1}+\frac{\alpha _{1}}{2}N\left\vert
\Psi \right\vert ^{2})\Psi & =\mu _{1}\Psi \\
(\frac{\hbar ^{2}}{2M_{2}}\nabla ^{2}+U_{2}+\frac{\alpha _{2}}{2}N\left\vert
\Phi \right\vert ^{2})\Phi & =\mu _{2}\Phi  \notag
\end{align}%
here $\mu _{1,2}$ is the mean field energy or the chemical potential in the
two wells.

The spin-dependent Hamiltonian finely reduce to
\begin{equation}
\hat{H}_{1}=C_{1}\mathbf{\hat{F}}_{1}^{2},\text{ }\hat{H}_{2}=C_{2}\mathbf{%
\hat{F}}_{2}^{2},
\end{equation}
with $C_{1}=\beta _{1}\int d\mathbf{r}\left\vert \Psi (r)\right\vert
^{4},C_{2}=\beta _{2}\int d\mathbf{r}\left\vert \Phi (r)\right\vert ^{4}$.

In the one-dimensional double well or optical lattice, according to the the
vector subtraction, $\mathbf{r}_{12}=\mathbf{r}_{1}-\mathbf{r}_{2}$ always
point at the only one direction. If we choose the quantization axis along
this direction as z axis, the $V_{dd}$ reduced to

\begin{eqnarray}
V_{dd} &=&\lambda \left[ \mathbf{F}_{1}\cdot \mathbf{F}_{2}-3(\mathbf{F}%
_{1}\cdot \mathbf{\hat{r}}_{12})(\mathbf{F}_{2}\cdot \mathbf{\hat{r}}_{12})%
\right]  \notag \\
&=&\lambda \lbrack \mathbf{F}_{1}\cdot \mathbf{F}_{2}-3\mathbf{F}_{1z}%
\mathbf{F}_{2z}]
\end{eqnarray}%
with $\lambda =\frac{\mu _{0}(g_{F}\mu _{B})^{2}}{4\pi \left\vert \mathbf{r}%
_{12}\right\vert ^{3}}.$The Hamiltonian $\hat{H}_{12}$ finely reads
\begin{equation}
\hat{H}_{12}=\Lambda (\mathbf{\hat{F}}_{1}\cdot \mathbf{\hat{F}}_{2}-3\hat{F}%
_{1z}\hat{F}_{2z})
\end{equation}%
with $\Lambda =\lambda \int d\mathbf{r}\left\vert \Psi (r)\right\vert
^{2}\left\vert \Phi (r)\right\vert ^{2}.$

The total Hamiltonian \cite{dipole4,dipole5} is
\begin{equation}
\hat{H}=C_{1}\mathbf{\hat{F}}_{1}^{2}+C_{2}\mathbf{\hat{F}}_{2}^{2}+\Lambda
\mathbf{\hat{F}}_{1}\cdot \mathbf{\hat{F}}_{2}-3\Lambda \hat{F}_{1z}\hat{F}%
_{2z}  \label{HHH}
\end{equation}%
In the absence of long-range magnetic dipole-dipole interaction or external
magnetic fields, there is no spin correlations between sites. For the Rb$%
^{87}$ condensate, the ground state in the individual sites favors
polarizing all the spins to the same direction, therefore they can be
considered as independent \textquotedblleft magnets\textquotedblright\ whose
pseudospin vectors point in random directions. But for individual Na$^{23}$
condensate, they favors vanishing the total spin in each site. As $\eta =\mu
_{0}(g_{F}\mu _{B})^{2}/4\pi \left\vert \mathbf{r}_{12}\right\vert ^{3}$ can
be greatly enhanced by the light-induced optical dipolar interaction if one
chooses appropriate laser fields to form the potential well \cite%
{ehanced1,ehanced2}$,$we aim to determine the spin structure of the system
if the different sites are allowed to interact with each other through the
magnetic dipole-dipole interaction.

\section{The ground state properties}

\subsection{A brief review of singlet state}

Without the magnetic dipole-dipole interaction or for the intra-site pure
spin-1 condensate $(\hat{H}_{0}=C\mathbf{\hat{F}}^{2})$, the simplest ground
state for the F=1 spinor $^{23}$Na condensates $(C>0)$ is a spin singlet
formed by two spin-1 particles described as
\begin{equation}
\left\vert F,m\right\rangle =\underset{m_{1},m_{2}=1,0,-1}{\sum }G\left\vert
F_{1}=1,m_{1}\right\rangle \left\vert F_{2}=1,m_{2}\right\rangle
\end{equation}%
with $F=F_{1}+F_{2}=0$ is the total spin, $m=m_{1}+m_{2}=0$ is the total z
component, $G$ is the Clebsch-Gordon coefficient.

\begin{equation}
\left\vert F,m\right\rangle =\frac{1}{\sqrt{3}}\hat{A}^{\dag }\left\vert
0\right\rangle
\end{equation}%
The operator $\hat{A}^{\dag }\equiv (\hat{a}_{0}^{\dag })^{2}-2\hat{a}%
_{1}^{\dag }\hat{a}_{-1}^{\dag }$ describe a singlet pair creating operator
formed by two identical spin-1 bonsons, and the ground state of N particles
is $(\hat{A}^{\dag })^{N/2}\left\vert 0\right\rangle .$ G is the CG
coefficient. The particle density matrix $(\hat{\rho})_{\alpha \beta
}=\left\langle \hat{a}_{\alpha }^{\dag }\hat{a}_{\beta }\right\rangle $ is

\begin{equation}
\left\langle \hat{a}_{\alpha }^{\dag }\hat{a}_{\beta }\right\rangle =\left(
\begin{array}{ccc}
N/3 &  &  \\
& N/3 &  \\
&  & N/3%
\end{array}%
\right)
\end{equation}%
with $\alpha ,\beta $=$1,0,-1.$This matrix has three equal macroscopic
eigenvalues called \textquotedblleft superfragmented state\textquotedblright
\cite{Ho and Yip}. A weak external magnetic field along z\ $(\hat{H}_{0}=C%
\mathbf{\hat{F}}^{2}-p\hat{F}_{z})$ can break the pairs and polarize the
system \cite{fragment1,fragment2} with the ground state described as%
\begin{equation}
\left\vert F=S,m=S\right\rangle =(\hat{a}_{1}^{\dag })^{S}(\hat{A}^{\dag
})^{\left( N-S\right) /2}\left\vert 0\right\rangle  \label{fff}
\end{equation}%
The particle numbers on the three Zeeman levels are redistributed as

\begin{eqnarray*}
N_{1} &=&\frac{\left( N+S\right) (S+1)}{2S+3}+\frac{S}{2S+3} \\
N_{-1} &=&\frac{\left( N-S\right) (S+1)}{2S+3} \\
N_{0} &=&\frac{N-S}{2S+3}
\end{eqnarray*}
with the 0-component distribution shrink rapidly as S increases.

\subsection{A brief review of dimmer state}

For the subspace of exactly one particle per well, the Mott-insulator ground
state of one-dimensional optical lattice has been confirmed to be a dimmer
states\cite{Yip} with the form,

\begin{equation}
\Psi _{dimer}=\Psi _{12}\Psi _{34}\Psi _{56}...
\end{equation}%
where the state%
\begin{eqnarray}
\Psi _{12} &=&\frac{-1}{\sqrt{3}}(\left\vert 1,-1\right\rangle +\left\vert
-1,1\right\rangle -\left\vert 0,0\right\rangle )_{12}  \notag \\
&=&\frac{1}{\sqrt{3}}(\hat{a}_{0}^{\dag }\hat{b}_{0}^{\dag }-\hat{a}%
_{1}^{\dag }\hat{b}_{-1}^{\dag }-\hat{a}_{-1}^{\dag }\hat{b}_{1}^{\dag
})\left\vert 0\right\rangle  \notag \\
&=&\frac{1}{\sqrt{3}}\hat{\Theta}_{12}^{\dag }\left\vert 0\right\rangle
\end{eqnarray}%
The operator $\hat{\Theta}_{12}^{\dag }\equiv \hat{a}_{0}^{\dag }\hat{b}%
_{0}^{\dag }-\hat{a}_{1}^{\dag }\hat{b}_{-1}^{\dag }-\hat{a}_{-1}^{\dag }%
\hat{b}_{1}^{\dag }$ describe a singlet pair creating operator formed by two
spin-1 bonsons in the different site.

For more or at least two particles per well, the ground states are more
complicated.

\subsection{The quaternary-singlet State}

For simplify, we first consider $3\Lambda \hat{F}_{1z}\hat{F}_{2z}=0$, which
serves as reference case for the complete discussion.

\begin{equation}
\hat{H}=C_{1}\mathbf{\hat{F}}_{1}^{2}+C_{2}\mathbf{\hat{F}}_{2}^{2}+\Lambda
\mathbf{\hat{F}}_{1}\cdot \mathbf{\hat{F}}_{2}  \label{Ham2}
\end{equation}%
It can be rewrited as\cite{lizhibing,Xuone,Xutwo,zj,zj2,shiyu,shiyu2}
\begin{equation}
\hat{H}=a\mathbf{\hat{F}}_{1}^{2}+b\mathbf{\hat{F}}_{2}^{2}+c\mathbf{\hat{F}}%
^{2},  \label{mH}
\end{equation}%
with $a=C_{1}-\Lambda /2$, $b=C_{2}-\Lambda /2$, and $c=\Lambda /2,\mathbf{%
\hat{F}}=\mathbf{\hat{F}}_{1}+\mathbf{\hat{F}}_{2}$ is the total spin
operator. The eigenstates of (\ref{mH}) are the common eigenstates for the
commuting operators $\mathbf{\hat{F}}_{1}^{2},\mathbf{\hat{F}}_{2}^{2},%
\mathbf{\hat{F}}^{2}$, and $\hat{F}_{z}$, given by
\begin{equation}
\left\vert F_{1},F_{2},F,m\right\rangle
=\sum_{m_{1}m_{2}}C_{F_{1,}m_{1};F_{2,}m_{2}}^{F,m}\left\vert
F_{1},m_{1}\right\rangle \left\vert F_{2},m_{2}\right\rangle ,  \label{basis}
\end{equation}%
with the uncoupled basis states $\left\vert F_{1},m_{1}\right\rangle $ ($%
\left\vert F_{2},m_{2}\right\rangle )$ generated from equation (\ref{fff})
by a repeat using lowering operator $\hat{F}_{1-},$ and they can span a
Hilbert space of dimension $(N_{1}+1)(N_{1}+2)/2$ \cite{MK}. $C$ is the
Clebsch-Gordon coefficient. The corresponding eigenenergy is
\begin{equation}
E=aF_{1}(F_{1}+1)+bF_{2}(F_{2}+1)+cF(F+1)  \label{Eval}
\end{equation}%
Given $N_{j}$, the allowed values of $F_{j}$ are $F_{j}=0,2,4,\cdots N_{j}$
if $N_{j}$ is even; and $F_{j}=1,3,5,\cdots N_{j}$ if $N_{j}$ is odd,
satisfying $\left\vert F_{1}-F_{2}\right\vert \leqslant F\leqslant
F_{1}+F_{2}$.

We will next consider the special case of $N_{1}=N_{2}=N$ and for $N$ even.

Fig. 1 shows the development of the four order parameters%
\begin{eqnarray}
\mathbf{\bar{F}}_{1}^{2} &=&<\mathbf{\hat{F}}_{1}^{2}>,\text{ }  \notag \\
\mathbf{\bar{F}}_{2}^{2} &=&<\mathbf{\hat{F}}_{2}^{2}>,  \notag \\
\mathbf{\bar{F}}_{1}\cdot \mathbf{\bar{F}}_{2} &=&<\mathbf{\hat{F}}_{1}\cdot
\mathbf{\hat{F}}_{2}>,  \notag \\
\mathbf{\bar{F}}^{2} &=&<\mathbf{\hat{F}}^{2}>
\end{eqnarray}%
for $\Lambda $ (in the unit of $\left\vert C_{1}\right\vert $) with even
particles per well ( take $N_{1}=N_{2}=6$).We find that in the $\Lambda <%
\frac{-({2N-1})C_{2}}{C_{1}N}$ region, $\mathbf{\bar{F}}_{1}^{2},$ $\mathbf{%
\bar{F}}_{2}^{2},$ and $\mathbf{\bar{F}}_{1}\cdot \mathbf{\bar{F}}_{2}$ are
all polarized to the maximum with the system being ferromagnetic. In the
region $\Lambda \in \left[ -C_{1}-C_{2},C_{1}+C_{2}\right] $, the two sites
are essentially independent for a weak inter-sites dipole-dipole
interaction. This phase is a total spin singlet $\mathbf{\bar{F}}^{2}=0$
described by the direct product of the polar ground state $(\hat{A}^{\dag
})^{N/2}(\hat{B}^{\dag })^{N/2}\left\vert 0\right\rangle $ giving rise to $%
\mathbf{\bar{F}}_{1}^{2}=0$ $\mathbf{\bar{F}}_{2}^{2}=0$and $\mathbf{\bar{F}}%
_{1}\cdot \mathbf{\bar{F}}_{2}=0.$When $\Lambda >\frac{({2N-1})C_{2}}{C_{1}({%
N+1})}$ they are polarized to the maximum but in the opposite directions
with $\mathbf{\bar{F}}^{2}=0$ and $-2\mathbf{\bar{F}}_{1}\cdot \mathbf{\bar{F%
}}_{2}=\mathbf{\bar{F}}_{1}^{2}+\mathbf{\bar{F}}_{2}^{2}\neq 0$.We find
interestingly that in this state the total spin vanishes, while the sites
spins satisfy $\mathbf{\bar{F}}_{1}^{2}=\mathbf{\bar{F}}_{2}^{2}=N(N+1)$ .

\begin{figure}[tbp]
\includegraphics[width=3.0in]{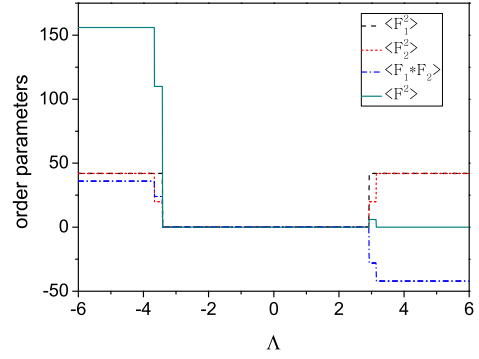}
\caption{(Color online) The dependence of ground-state order parameters on $%
\Lambda $ at fixed values of $C_{1}=1,$ $C_{2}=2,$ (in the unit of $%
\left\vert C_{1}\right\vert $). Black dashed lines, red short dashed lines,
blue dot-dashed lines and green solid lines denote respectively the order
parameters $\mathbf{\bar{F}}_{1}^{2},\mathbf{\bar{F}}_{2}^{2},\mathbf{\bar{F}%
}_{1}\cdot \mathbf{\bar{F}}_{2}$, and $\mathbf{\bar{F}}^{2}.$}
\label{2002}
\end{figure}

The ground state in the region $\Lambda >\frac{({2N-1})C_{2}}{C_{1}({N+1})}$
is a singlet, with all basis states obeying the condition $m_{1}+m_{2}=0$.
All channels of total spin zero have to be taken

into account and we have
\begin{equation}
\left\vert N,N,0,0\right\rangle =\underset{m_{1}=-N}{\overset{N}{\sum }}%
C_{N,m_{1};N,-m_{1}}^{0,0}\left\vert N,m_{1}\right\rangle \left\vert
N,-m_{1}\right\rangle .\hskip12pt  \label{AA}
\end{equation}

If we take $N_{1}$=$N_{2}$=$2$ for example, we find that
\begin{equation}
\left\vert 2,2,0,0\right\rangle =\frac{1}{2\sqrt{5}}((\hat{\Theta}%
_{12}^{\dag })^{2}-\frac{1}{3}\hat{A}^{\dag }\hat{B}^{\dag })\left\vert
0\right\rangle
\end{equation}
Compared to the \textquotedblleft superfragmented state\textquotedblright\ $(%
\hat{A}^{\dag })^{N/2}\left\vert 0\right\rangle =((\hat{a}_{0}^{\dag })^{2}-2%
\hat{a}_{1}^{\dag }\hat{a}_{-1}^{\dag })^{N/2}\left\vert 0\right\rangle $%
\cite{Ho and Yip}, this quaternary-singlet states has the similar
appearance. However it is not the direct product of the two conventional
singlet pairs, $\left\vert 2,2,0,0\right\rangle \neq \hat{A}^{\dag }\hat{B}%
^{\dag }\left\vert 0\right\rangle .$ The difference between $\left\vert
N,N,0,0\right\rangle $ and $(\hat{A}^{\dag })^{N/2}(\hat{B}^{\dag
})^{N/2}\left\vert 0\right\rangle $ can be easily find in the phase diagram
(Fig.1).

\subsection{The number fluctuations}

\bigskip We notice that
\begin{eqnarray}
\lbrack \mathbf{\hat{F}}_{1}^{2},\hat{\Theta}_{12}^{\dag }\hat{\Theta}_{12}]
&\neq &0,\ [\mathbf{\hat{F}}_{2}^{2},\hat{\Theta}_{12}^{\dag }\hat{\Theta}%
_{12}]\neq 0,  \notag \\
\lbrack \mathbf{\hat{F}}^{2},\hat{\Theta}_{12}^{\dag }\hat{\Theta}_{12}] &=&0
\end{eqnarray}%
and it is easy to understand that $\mathbf{\hat{F}}^{2}(\hat{\Theta}%
_{12}^{\dag })^{N}\left\vert 0\right\rangle =0.$ In this section, we will
talk about three different singlet states, the quaternary-singlet\ state $%
\left\vert N,N,0,0\right\rangle ,$ the dimer state $(\hat{\Theta}_{12}^{\dag
})^{N}\left\vert 0\right\rangle $ and the direct product singlet state $(%
\hat{A}^{\dag })^{N/2}(\hat{B}^{\dag })^{N/2}\left\vert 0\right\rangle ,$
the properties such as
\begin{equation}
\mathbf{\hat{F}}^{2}(\hat{A}^{+})^{N_{1}}(\hat{B}^{+})^{N_{2}}\left\vert
0\right\rangle =0,\mathbf{\hat{F}}^{2}\left\vert N,N,0,0\right\rangle =0
\end{equation}%
are shown in Fig.1.

As a exceptional case, all eigenvectors\ $\left\vert N,m_{1}\right\rangle $
in (\ref{basis}) can be expressed in terms of the Fock states \cite{Ying Wu}%
, which are defined as
\begin{eqnarray}
\hat{n}_{\alpha }^{(j)}\left\vert
n_{1}^{(j)},n_{0}^{(j)},n_{-1}^{(j)}\right\rangle &=&n_{\alpha
}^{(j)}\left\vert n_{1}^{(j)},n_{0}^{(j)},n_{-1}^{(j)}\right\rangle ,  \notag
\\
\alpha &=&0,\pm 1;j=1,2
\end{eqnarray}%
For the state $\left\vert N,N,0,0\right\rangle ,$we calculate the particle
numbers and number fluctuations on the Fock states and find that the average
numbers of atoms in the six components are exactly all equal, $\left\langle
n_{1}^{(j)}\right\rangle =\left\langle n_{0}^{(j)}\right\rangle
=\left\langle n_{-1}^{(j)}\right\rangle =N/3$. The fluctuations are given
explicitly
\begin{eqnarray}
\left\langle \Delta n_{0}^{(j)}\right\rangle &=&\frac{\sqrt{N^{2}+9N}}{3%
\sqrt{5}}  \notag \\
\left\langle \Delta n_{\pm 1}^{(j)}\right\rangle &=&\frac{2\sqrt{N^{2}+3N/2}%
}{3\sqrt{5}}  \label{flu}
\end{eqnarray}%
which approximatively satisfy $\left\langle \Delta n_{1}^{(j)}\right\rangle
=2\left\langle \Delta n_{0}^{(j)}\right\rangle =\left\langle \Delta
n_{-1}^{(j)}\right\rangle $ for large $N$\cite{zj},
as opposed to $2\left\langle \Delta n_{1}\right\rangle =\left\langle \Delta
n_{0}\right\rangle =2\left\langle \Delta n_{-1}\right\rangle $ for the
single species or intra-site singlet state $(C_{1,2}>0)$ \cite{Ho and Yip}.

The difference between these states are obvious. As total spin F vanishes,
the number distributions are all $\left\langle n_{1}^{(j)}\right\rangle
=\left\langle n_{0}^{(j)}\right\rangle =\left\langle
n_{-1}^{(j)}\right\rangle =N/3,$ but the number fluctuation distribution in
these states are quite different, it has been shown that for the state $%
Z^{1/2}(\hat{A}^{+})^{N_{1}}(\hat{B}^{+})^{N_{2}}\left\vert 0\right\rangle $
\cite{Ho and Yip}, they are
\begin{eqnarray}
\left\langle \Delta n_{1}^{(j)}\right\rangle  &=&\left\langle \Delta
n_{0}^{(j)}\right\rangle /2=\left\langle \Delta n_{-1}^{(j)}\right\rangle
\notag \\
&=&\frac{\sqrt{N^{2}+3N}}{3\sqrt{5}}
\end{eqnarray}%
For the state $Z^{1/2}(\hat{\Theta}_{12}^{\dag })^{N}\left\vert
0\right\rangle ,$ according to the multinomial theorem
\begin{equation}
\left( x_{1}+x_{2}+x_{3}\right)
^{n}=\sum_{k=0}^{n}\sum_{l=0}^{k}c_{nlk}x_{1}^{n-k}x_{2}^{k-l}x_{3}^{l}
\end{equation}%
with $c_{nlk}=n!/\left( l!(k-l)!(n-k)!\right) $, we find that the state $(%
\hat{\Theta}_{12}^{\dag })^{N}\left\vert 0\right\rangle $ can be described
by the Fock state $\left\vert
n_{1}^{(1)},n_{0}^{(1)},n_{-1}^{(1)}\right\rangle \otimes \left\vert
n_{1}^{(2)},n_{0}^{(2)},n_{-1}^{(2)}\right\rangle $ as%
\begin{eqnarray}
&&(\hat{\Theta}_{12}^{\dag })^{N}\left\vert 0\right\rangle =\left( \hat{a}%
_{0}^{\dag }\hat{b}_{0}^{\dag }-\hat{a}_{1}^{\dag }\hat{b}_{-1}^{\dag }-\hat{%
a}_{-1}^{\dag }\hat{b}_{1}^{\dag }\right) ^{N}\left\vert 0\right\rangle
\notag \\
&=&\sum_{k=0}^{N}\sum_{l=0}^{k}c_{Nlk}(\hat{a}_{0}^{\dag }\hat{b}_{0}^{\dag
})^{N-k}(-\hat{a}_{1}^{\dag }\hat{b}_{-1}^{\dag })^{k-l}(-\hat{a}_{-1}^{\dag
}\hat{b}_{1}^{\dag })^{l}\left\vert 0\right\rangle   \notag \\
&=&\sum_{k=0}^{N}\sum_{l=0}^{k}(-1)^{k}N!\left\vert k-l,N-k,l\right\rangle
\otimes \left\vert l,N-k,k-l\right\rangle \hskip18pt
\end{eqnarray}%
where we have used the property $\left( \hat{a}^{\dag }\right)
^{N}\left\vert 0\right\rangle =\sqrt{N!}\left\vert N\right\rangle $. We find
that the number fluctuations are equally distributed, i.e.
\begin{eqnarray}
\left\langle \Delta n_{1}^{(j)}\right\rangle  &=&\left\langle \Delta
n_{0}^{(j)}\right\rangle =\left\langle \Delta n_{-1}^{(j)}\right\rangle
\notag \\
&=&\sqrt{N(N+1)/6-N^{2}/9}
\end{eqnarray}

\subsection{The term $3\Lambda \hat{F}_{1z}\hat{F}_{2z}$}

\bigskip For the real Hamiltonian (\ref{HHH}), the last term plays an
important role in domain formation and can polarize the spin to the same
direction. This interaction offers a effect extra uniform weak field to the
nearby site and breaks the singlet states. The ground state can be
constructed using the quaternary-singlet\ state $\left\vert
N,N,0,0\right\rangle $ and direct product singlet state $(\hat{A}^{\dag
})^{N/2}(\hat{B}^{\dag })^{N/2}\left\vert 0\right\rangle .$

In the region $\Lambda \in \left[ -C_{1}-C_{2},C_{1}+C_{2}\right] $, since
the spin singlet operator commutes with the spin $\ \ $

\begin{equation}
\lbrack \mathbf{\hat{F}}_{1}^{2},\hat{A}^{\dag }]=0,\ [\mathbf{\hat{F}}%
_{2}^{2},\hat{B}^{\dag }]=0
\end{equation}%
and it does not change total spin and any spin components but just add two
particles. Therefore, we can construct the unnormalized spin state for N
particles\cite{before2}: first, write down a state with necessary spin for a
small number of particles; second, apply $\hat{A}^{\dag }(\hat{B}^{\dag })$
as many times as needed to get the desired number of particles. We got
\begin{equation}
\left\vert \otimes \right\rangle =Z^{1/2}(\hat{a}_{1}^{\dag })^{S_{1}}(\hat{b%
}_{1}^{\dag })^{S_{2}}(\hat{A}^{\dag })^{T_{1}}(\hat{B}^{\dag
})^{T_{2}}\left\vert 0\right\rangle
\end{equation}%
with the fixed number in the two sites satisfied $%
T_{1,2}=(N_{1,2}-S_{1,2})/2.$

For the region $\Lambda >\frac{({2N-1})C_{2}}{C_{1}({N+1})},$ if we let $%
\left\vert N,N,0,0\right\rangle =Z^{1/2}(\hat{Q}^{\dag })^{2N}\left\vert
0\right\rangle ,$the ground state for the Hamiltonian (\ref{HHH}) is
\begin{equation}
\left\vert Q\right\rangle =Z^{1/2}(\hat{a}_{1}^{\dag })^{S}(\hat{b}%
_{1}^{\dag })^{S}(\hat{Q}^{\dag })^{2N-2S}\left\vert 0\right\rangle
\end{equation}

In the $\Lambda <\frac{-({2N-1})C_{2}}{C_{1}N}$ region, the system is
polarized to the ferromagnetic phase, the ground state is
\begin{equation}
\left\vert P\right\rangle =Z^{1/2}(\hat{a}_{1}^{\dag })^{N_{1}}(\hat{b}%
_{1}^{\dag })^{N_{2}}\left\vert 0\right\rangle .
\end{equation}

\section{Conclusion}

To summarize, we study the ground spin state of polar atoms ($^{23}$Na) in
the optical lattice subject to a magnetic dipole-dipole interaction between
nearby wells. We consider the special case that there are two particles per
well, and show a new singlet state. In two well model, three kinds of spin
ground state with total spin vanished $(\mathbf{\bar{F}}^{2}=\left\langle (%
\mathbf{\hat{F}}_{1}\cdot \mathbf{\hat{F}}_{2})^{2}\right\rangle =0)\ $are
discussed and can be distinguished by the number fluctuations. The final
states can be constructed by singlet pair creation operator and the
quaternary-singlet creation operator.


\bigskip

\begin{thebibliography}{99}
\bibitem{D. M. Stamper-Kurn} D. M. Stamper-Kurn, M. R. Andrews, A. P.
Chikkatur, S. Inouye, H.-J. Miesner, J. Stenger, and W. Ketterle, Phys. Rev.
Lett. \textbf{80}, 2027 (1998).

\bibitem{Stenger} J. Stenger et al., Nature (London) \textbf{396}, 345
(1998).

\bibitem{MSChang} M.-S. Chang, C. D. Hamley, M. D. Barrett, J. A. Sauer,
K.M. Fortier, W. Zhang, L. You, and M. S. Chapman, Phys. Rev. Lett. \textbf{%
92}, 140403 (2004).

\bibitem{Ho} Tin-Lun Ho, Phys. Rev. Lett. \textbf{81}, 742 (1998).

\bibitem{Ohmi} T. Ohmi and K. Machida, J. Phys. Soc. Jpn. \textbf{67}, 1822
(1998).

\bibitem{BH} D. Jaksch et al., Phys. Rev. Lett. 81, 3108 (1998).

\bibitem{BH2} M.P.A. Fisher et al., Phys. Rev. B 40, (1989).

\bibitem{BHEX} C. Orzel et al., Science 291, 2386 (2001).

\bibitem{BHEX2} M. Greiner et al., Nature (London) 415, 39 (2002).

\bibitem{before1} E. Demler and F. Zhou, Phys. Rev. Lett. 88, 163001 (2002).

\bibitem{before2} A. Imambekov, M. Lukin, and E. Demler, Phys. Rev. A 68,
063602 (2003).

\bibitem{before3} S.K. Yip, Phys. Rev. Lett. 90, 250402 (2003).

\bibitem{before4} S.K. Yip, J. Phys. B: Cond. Mat., 15, 4583 (2003).

\bibitem{before5} M. Rizzi, D. Rossini, G. De Chiara, S. Montangero, and R.
Fazio, Phys Rev. Lett. 95, 240404 (2005). ).

\bibitem{before6} F. Zhou and G.W. Semenoff, Phys. Rev. Lett. 97, 180411
(2006).

\bibitem{before7} New Journal of Physics 9 (2007) 133.

\bibitem{dipole1} K. Goral et al., Phys. Rev. A 61, 051601 (2000).

\bibitem{dipole2} L. Santos et al., Phys. Rev. Lett. 85, 1791 (2000).

\bibitem{dipole3} S. Yi and L. You, Phys. Rev. A 63, 053607 (2001).

\bibitem{dipole4} H. Pu, W. Zhang, and P.Meystre, Phys. Rev. Lett. 87,
140405 (2001).

\bibitem{dipole5} H. Pu, W. Zhang, and P.Meystre, Phys. Rev. Lett. 89,
090401 (2002).

\bibitem{dipole6} S. Yi and H. Pu, Phys. Rev. Lett. 97, 020401 (2006); S. Yi
and H. Pu, Phys. Rev. A 73, 061602(R) (2006); S. Yi and H. Pu, Phys. Rev. A
73, 023602 (2006).

\bibitem{dipole7} B. Sun, W. X. Zhang, S. Yi, M. S. Chapman, and L. You,
Phys. Rev. Lett. 97, 123201 (2006); Phys. Rev. Lett. 97, 139902 (2006).

\bibitem{dipole8} S. Yi, T. Li, and C. P. Sun, Phys. Rev. Lett. 98, 260405
(2007).

\bibitem{C. K. Law} C. K. Law, H. Pu, and N. P. Bigelow, Phys. Rev. Lett.
\textbf{81}, 5257 (1998).

\bibitem{Ho and Yip} T.-L. Ho and S.-K. Yip, Phys. Rev. Lett. \textbf{84},
4031 (2000).

\bibitem{SMA1} H. Pu, C. K. Law, S. Raghavan, J. H. Eberly, and N. P.
Bigelow, Phys. Rev. A 60, 1463 (1999); H. Pu, S. Raghavan, and N. P.
Bigelow, ibid. 61, 023602 (2000).

\bibitem{SMA2} S. Yi, O. E. Mu,tecaplioglu, C. P. Sun, and L. You, Phys.
Rev. A 66, 011601(R) (2002).

\bibitem{ehanced1} W. Zhang, H. Pu, C. Search, and P. Meystre, Phys. Rev.
Lett. 88, 060401 (2002).

\bibitem{ehanced2} S. Giovanazzi, D. O'Dell, and G. Kurizki, Phys. Rev.
Lett. 88, 130402 (2002).

\bibitem{fragment1} M. Koashi and M. Ueda, Phys. Rev. Lett. 84, 1066 (2000).

\bibitem{fragment2} M. Ueda and M. Koashi, Phys. Rev. A 65, 063602 (2002).

\bibitem{lizhibing} M. Luo, Z. B. Li, and C. G. Bao, Phys. Rev. A 75, 043609
(2007).

\bibitem{Xuone} Z. F. Xu, Y. Zhang, and L. You, Phys. Rev. A \textbf{79},
023613 (2009).

\bibitem{Xutwo} Z. F. Xu, J. Zhang, Y. Zhang, and L. You, Phys. Rev. A
\textbf{81}, 033603 (2010).

\bibitem{zj} J. Zhang, Z. F. Xu, L. You and Y. Zhang, Phys. Rev. A \textbf{82%
}, 013625 (2010).

\bibitem{zj2} J. Zhang, T. T. Li,Y. Zhang, Phys. Rev. A \textbf{83}, 023614
(2011).

\bibitem{shiyu} Y. Shi, Phys. Rev. A \textbf{82}, 023603 (2010).

\bibitem{shiyu2} Y. Shi and L. Ge, Phys. Rev. A \textbf{83}, 013616 (2011).

\bibitem{Yip} S. K. Yip, Phys. Rev. Lett. \textbf{90}, 250402 (2003).

\bibitem{MK} Masato Koashi and Masahito Ueda, Phys. Rev. Lett. \textbf{84},
1066 (2000).

\bibitem{Ying Wu} Ying Wu, Phys. Rev. A \textbf{54}, 4534 (1996).
\end{thebibliography}
\end{document}